\documentclass{jetpl}
\usepackage{graphicx}
\usepackage{amssymb}
\usepackage{amsmath}
\usepackage{color}
\usepackage{hyperref}
\twocolumn

\title{Oscillating nematic aerogel in superfluid $^3$He}
\rtitle{Oscillating nematic aerogel in superfluid $^3$He}
\sodtitle{Oscillating nematic aerogel in superfluid $^3$He}

\author{V.\,V.\,Dmitriev$^+$\thanks{e-mail: dmitriev@kapitza.ras.ru},\,M.\,S.\,Kutuzov$^*$,\,A.\,A.\,Soldatov$^+$,\,E.\,V.\,Surovtsev$^+$, and \,A.\,N.\,Yudin$^+$}
\rauthor{V.\,V.\,Dmitriev,\,M.\,S.\,Kutuzov,\,A.\,A.\,Soldatov,\,E.\,V.\,Surovtsev,\,A.\,N.\,Yudin}
\sodauthor{V.\,V.\,Dmitriev,\,M.\,S.\,Kutuzov,\,A.\,A.\,Soldatov,\,E.\,V.\,Surovtsev,\,A.\,N.\,Yudin}

\address{$^+$P.\,L. Kapitza Institute for Physical Problems of RAS, 119334 Moscow, Russia}
\address{$^*$Metallurg Engineering Ltd., 11415 Tallinn, Estonia}

\dates{\selectlanguage{english}\today}{*}

\begin{document}
\abstract{We present experiments on nematic aerogel oscillating in superfluid $^3$He. This aerogel consists of nearly parallel mullite strands and is attached to a vibrating wire moving along the direction of the strands. Previous nuclear magnetic resonance experiments in $^3$He confined in similar aerogel sample have shown that the superfluid transition of $^3$He in aerogel occurs into the polar phase and the transition temperature ($T_{ca}$) is only slightly suppressed with respect to the superfluid transition temperature of bulk $^3$He. In present experiments we observed a change in resonant properties of the vibrating wire at $T=T_{ca}$ and found that below $T_{ca}$ an additional resonance mode is excited which is coupled to the main resonance.}

\maketitle

\section{Introduction}
Superfluidity of $^3$He in aerogel can be investigated using a vibrating wire (VW) resonator immersed in liquid $^3$He with an aerogel sample attached to it. In this case an appearance of the superfluid fraction of $^3$He in aerogel influences resonant properties of the VW. Experiments with aerogel attached to the VW have been done previously only with silica aerogel \cite{bru00,bru01,bra07,bra08} where superfluid phases (A-like and B-like) have the same order parameters as A and B phases of bulk $^3$He. These experiments have allowed to estimate the temperature dependence of the superfluid fraction in A-like and B-like phases as well as to detect an influence of the superfluid flow on the texture of the order parameter in the A-like phase.

In this Letter, we present results of experiments with $^3$He in the so-called nematic aerogel using VW resonator.
The nematic aerogel is a highly porous structure consisting of strands with almost parallel orientation \cite{asad15}.
It has been established that in this case the superfluid transition occurs into a new phase (polar phase)
that does not exist either in bulk $^3$He or in $^3$He in silica aerogel \cite{dmit15}. The polar phase becomes favorable due to an essentially anisotropic scattering of $^3$He quasiparticles inside the aerogel \cite{AI,fom14,ik15,fom18}. This phase has a superfluid gap with line of zeroes in the plane perpendicular to the specific direction \cite{VW,Elt,Aut}. In nematic aerogel this direction coincides with the average direction of the strands, along which the mean free path of $^3$He quasiparticles is maximal \cite{AI}.

\section{Sample and methods}
We used a sample of mullite nematic aerogel which has a form of cuboid with a size along strands $\approx2.6$\,mm and characteristic transverse sizes $\sim3\times3$\,mm. It was cut from a larger piece of the original sample synthesized by Metallurg Engineering Ltd so that it has perfectly flat ends (the planes where strands begin and end): irregularities are about 100\,nm. The sample consists of nearly parallel mullite strands with diameters of $\leq14$\,nm (estimated from the scanning transmission electron microscope images) and has the overall density $\approx150$\,mg/cm$^3$. If we assume that the density of mullite is 3.1\,g/cm$^3$ then the porosity of the sample is 95.2\% and the average distance between the strands is 60\,nm. Similar mullite sample (which was cut from the same original piece and was placed in a separate cell of the same experimental chamber) has been used in nuclear magnetic resonance (NMR) experiments in $^3$He \cite{dmit19,dmit20} where it was found that the superfluid transition of $^3$He in this sample actually occurs into the polar phase and that the superfluid transition temperature ($T_{ca}$) is only slightly suppressed with respect to the transition temperature ($T_c$) of bulk $^3$He. It was also found that on further cooling the second order transition into the polar-distorted A phase (the PdA phase) occurs, and that effective mean free paths of $^3$He quasiparticles in directions parallel and transverse to the aerogel strands in the limit of zero temperature are $\approx900$\,nm and $\approx235$\,nm correspondingly.
\begin{figure}[t]
\center
\includegraphics[width=0.8\columnwidth]{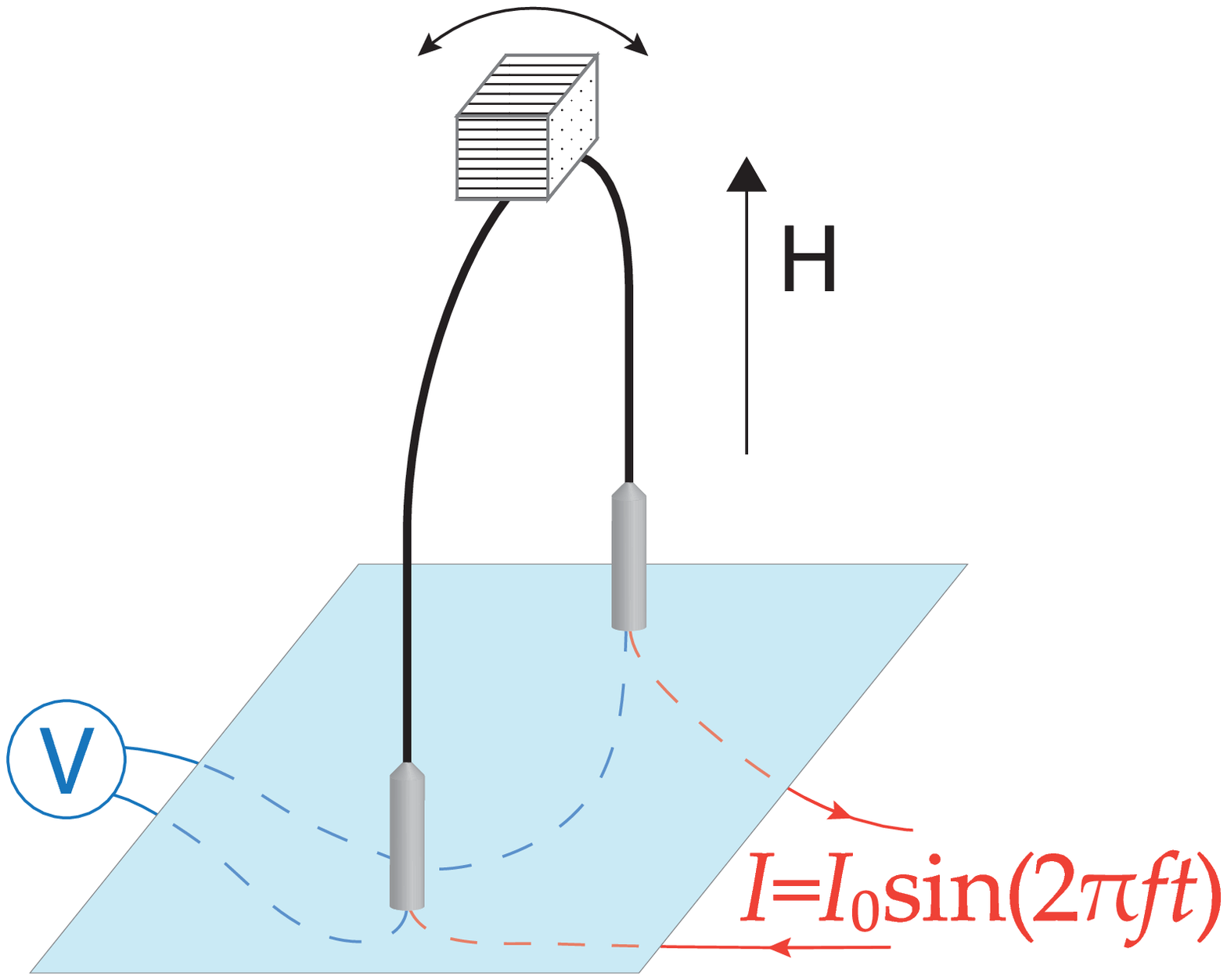}
\caption{Fig.\,\thefigure.
Signal measurement circuit of a VW immersed in liquid $^3$He in external steady magnetic field $\bf H$. The strands of nematic aerogel glued to the wire are oriented along the oscillations}
\label{wire}
\end{figure}

The present sample was glued using a small amount of Stycast-1266 epoxy resin to a 240\,$\mu$m NbTi wire,
bent into the shape of an arch with a total hight of 10\,mm and a distance between the legs of 8\,mm as shown in Fig.~\ref{wire}. The stycast was left to thicken until it was almost set before it was applied to the aerogel to prevent the aerogel from soaking up the glue. The wire is mounted in one of the cells of our experimental chamber. The experiments were carried out at pressures 7.1, 15.4, and 29.3\,bar and in magnetic fields 305--1650\,Oe. Necessary temperatures were obtained by a nuclear demagnetization cryostat and measured by a quartz tuning fork calibrated by Leggett frequency measurements in bulk $^3$He-B and $^3$He-A in separate NMR experiments. To stabilize the polar phase in nematic aerogel \cite{dmit18}, the samples were preplated with $\gtrsim2.5$ atomic layers of $^4$He.

A measurement procedure of the aerogel resonator is the same as in the case of a conventional wire resonator \cite{CHH}.
The mechanical flapping resonance of the wire is excited by the Lorentz force on an alternating current with amplitude $I_0$ (from 0.05\,mA to 0.5\,mA in our experiments), passing through the wire in a steady magnetic field. In liquid $^3$He the maximum velocity of the wire at such currents in the used range of temperatures did not exceed 0.1\,mm/s.
The motion of the wire generates a Faraday voltage which is amplified by a room-temperature step-up transformer {1:30} and measured with a lock-in amplifier. In-phase (dispersion) and quadrature (absorption) signals are joint fitted to Lorentz curves in order to extract a resonance frequency $f_a$ and a full width at half-maximum of the absorption (a resonance width) of the signal $\Delta f_a$. For our resonator, the resonance frequency in vacuum ($f_0$) is 752\,Hz.

\section{Theoretical model}
In $^3$He the resonance frequency of the VW resonator is inversely proportional to the square root of the effective
mass ($M$) which is oscillating. We use a simple model in which we neglect effects of $^3$He flow around the wire. We also do not consider effects due to a finite mean free path of $^3$He quasiparticles in bulk $^3$He because our measurements have been carried out at $T>0.6\,T_c$ where these effects can be neglected. Then $M$ has five contributions \cite{bra07,TMD,GWP,bla07}:
(i) the mass of the oscillating part of the wire and mass of the empty aerogel which sum ($m_0$) defines $f_0$, (ii) the mass of the normal-fluid fraction entrained in the aerogel ($m_n=\rho_n^a V$), (iii) the effective mass of the superfluid flow ($m_{sf}$), (iv) the effective mass of the normal component potential backflow ($m_{nf}=\alpha\rho_n V$), and (v) the effective mass $m_v$ which is carried by the body due to viscosity of the normal component of liquid $^3$He. Here $V$ is the volume of $^3$He in aerogel, $\rho_n^a$ and $\rho_n$ are the densities of normal components of $^3$He in aerogel and in bulk $^3$He, and $\alpha\sim1$ is a geometrical factor ($\alpha=0.5$ for a sphere and $\alpha=1$ for a cylinder oscillating along the direction normal to its axis). Then the expected resonant frequency equals
\begin{equation}\label{freqs}
f_a^2=f_0^2\frac{m_0}{m_0+m_n+m_{sf}+m_{nf}+m_v},
\end{equation}

The effective mass of the superfluid flow for an ellipsoidal sample moving along a principal axis can be found using analogy with a dielectric sample in an electric field \cite{GWP}:
\begin{equation}\label{ms}
m_{sf}=\alpha V\frac{(\rho_s-\rho_s^a)^2}{\rho_s+\alpha\rho_s^a},
\end{equation}
where $\rho_s^a$ and $\rho_s$ are the densities of superfluid components of $^3$He in aerogel and in bulk $^3$He.
It should be noted that Eq.~\eqref{ms} is obtained in the case of isotropic superfluid density tensors of the phases in the problem. The latter is valid only for the B phase of superfluid $^3$He. The superfluid density tensor of the polar phase (or the PdA phase) is anisotropic. As a result, the mass of the superfluid flow depends on the angle between the intrinsic anisotropy axis of aerogel and the direction of motion of the sample. In our case this angle is zero and $\rho_s^a$ is the superfluid density of the polar phase (or the PdA phase) along the strands.

The mass $m_v$ is related to the inertial part of the viscous drag force and can be estimated as $\rho_n S\delta$, where $\delta$ is the viscous penetration depth and $S$ is the surface area of the body. For a sphere with diameter $D$ the exact result is
\begin{equation}\label{mn}
m_v= \frac{3D^2}{4} \sqrt{\frac{\pi \rho_n \eta}{f}},
\end{equation}
where $\eta$ is the shear viscosity and $f$ is the frequency of oscillations \cite{LL}. The dissipative part of the viscous drag force determines the resonance width. If $\delta$ is much less than $D/2$ then in resonance $m_v\approx \rho_n V\Delta f_a/f_a$. Therefore, for the case of a general shape of the sample, the resonant frequency at $T>T_{ca}$ is expected to be given by
\begin{equation}\label{freq}
f_a^2=f_0^2\frac{m_0}{m_0+\rho V(1+\alpha)+\beta \rho_n V\Delta f_a/f_a},
\end{equation}
where $\beta$ is a geometrical factor ($\beta=1$ for a sphere) and $\rho$ is the density of $^3$He.
The resonant frequency $f_n$ in normal $^3$He in the limit of $\eta\rightarrow 0$ (that is $\Delta f_a\rightarrow 0$) satisfies the following condition:
\begin{equation}\label{fi}
\frac{1}{f_n^2}-\frac{1}{f_0^2}=\frac{(1+\alpha)V}{m_0 f_0^2}\rho.
\end{equation}
If $\Delta f_a \ll f_a$ then at $T>T_{ca}$
\begin{equation}\label{fw}
f_a=f_n-\frac{1}{2}\beta\Delta f_a.
\end{equation}

In the above reasoning, we have assumed that the normal and superfluid components of $^3$He
are separately incompressible. As it is shown in Ref.~\cite{CHH2} in superfluid $^3$He the compressibility affects Stokes parameters ($C$ and $C'$ in notations of Ref.~\cite{CHH2}) that affects $m_v$ (due to change of $C$) and $\Delta f_a$ (due to change of $C'$). Fortunately, these changes are not large and $C$ varies almost linearly with $C'$ \cite{CHH2}. Therefore, in the first approximation, in superfluid $^3$He $m_v$ also should be proportional to $\Delta f_a$.

\section{Experiments in normal $^3$He}
\begin{figure}[t]
\center
\includegraphics[width=\columnwidth]{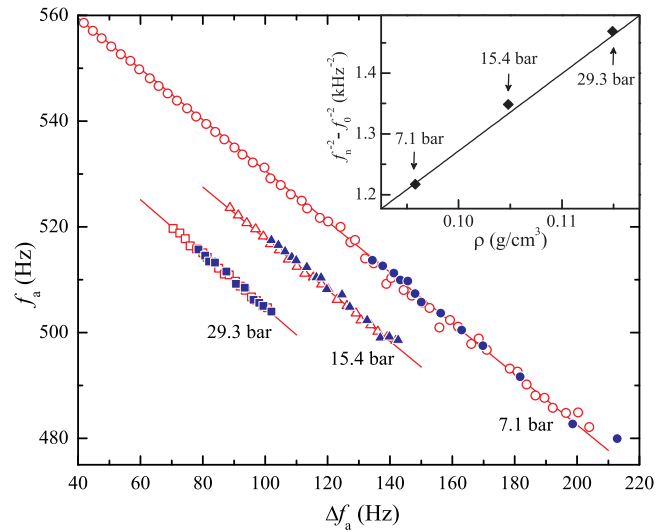}
\caption{Fig.\,\thefigure. 
The resonant frequency versus the resonance width measured at 29.3\,bar (circles), at 15.4\,bar (triangles), and at 7.1\,bar (squares). Open symbols correspond to measurements in normal $^3$He ($T>T_c$), filled symbols have been obtained at $T_{ca}<T<T_c$. Solid lines are fits to the data at $T>T_c$ using Eq.~\eqref{fw} (squares: $\beta=1.022$, triangles: $\beta=0.973$, circles: $\beta=0.962$). $I_0=0.25$\,mA, $H=1650$\,Oe. Inset: The dependence of $f_n$ on $\rho$ determined from linear fits shown in the main panel. Solid line is the best fit according to Eq.~\eqref{fi}.}
\label{fwn}
\end{figure}

In our experiments $\delta$ is essentially less than the characteristic sizes of the aerogel sample: even near $T_c$ at $P=29.3$\,bar $\delta\approx0.3$\,mm, so the observed resonance properties of our VW at $T>T_c$ are well described by the theoretical model. In Fig.~\ref{fwn} by open symbols we show dependencies of the resonant frequency on the resonance width measured at different pressures at $T>T_c$. It is seen that these dependencies agree with Eq.~\eqref{fw}. The solid lines in Fig.~\ref{fwn} are the best linear fits to the data which allow us to determine $f_n$ and $\beta$. We obtain that at all pressures $\beta$ is close to 1, that is to the value expected for a sphere. The obtained values of $f_n$ also agree with Eq.~\eqref{fi} (see the inset to Fig.~\ref{fwn}): the slope of the line in the inset is $12.7\times 10^{-6}$\,cm$^3$s$^2$/g while the value of the slope calculated from Eq.~\eqref{fi} (with $\alpha=0.5$ and estimated value of $m_0 \approx5.1$\,mg) is $11.6\times 10^{-6}$\,cm$^3$s$^2$/g.
We note that the dependence of $f_a$ on $\Delta f_a$ remains the same also at $T_{ca}<T<T_c$ (filled symbols in Fig.~\ref{fwn}). It means that the influence of compressibility of normal and superfluid components is not essential. Our experience with bare VWs resonators show that this dependence follows Eq.~\eqref{fw} down to $T\sim0.6\,T_c$.

\section{Experiments in superfluid $^3$He}
\begin{figure}[t]
\center
\includegraphics[width=\columnwidth]{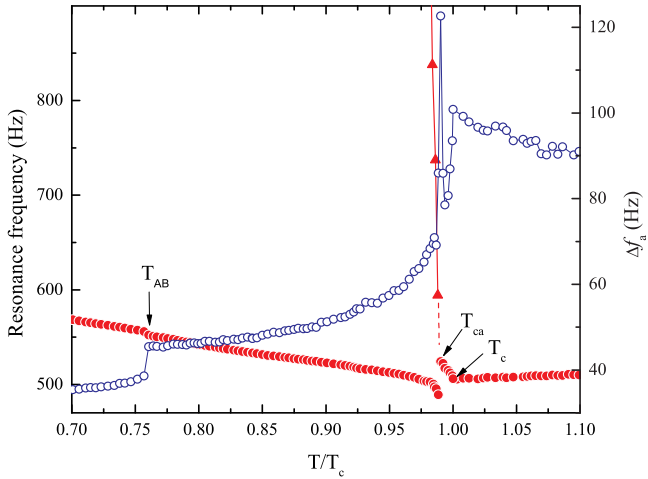}
\caption{Fig.\,\thefigure. 
Temperature dependencies of the resonance width of the main resonance (open circles) and of the frequencies of the main (filled circles) and the second (filled triangles) resonances. $P=29.3$\,bar, $I_0=0.25$\,mA, $H=1650$\,Oe. Arrows mark $T_{ca}$, $T_c$, and AB transition in bulk $^3$He at $T=T_{AB}$.}
\label{mode1}
\end{figure}

\begin{figure}[t]
\center
\includegraphics[width=\columnwidth]{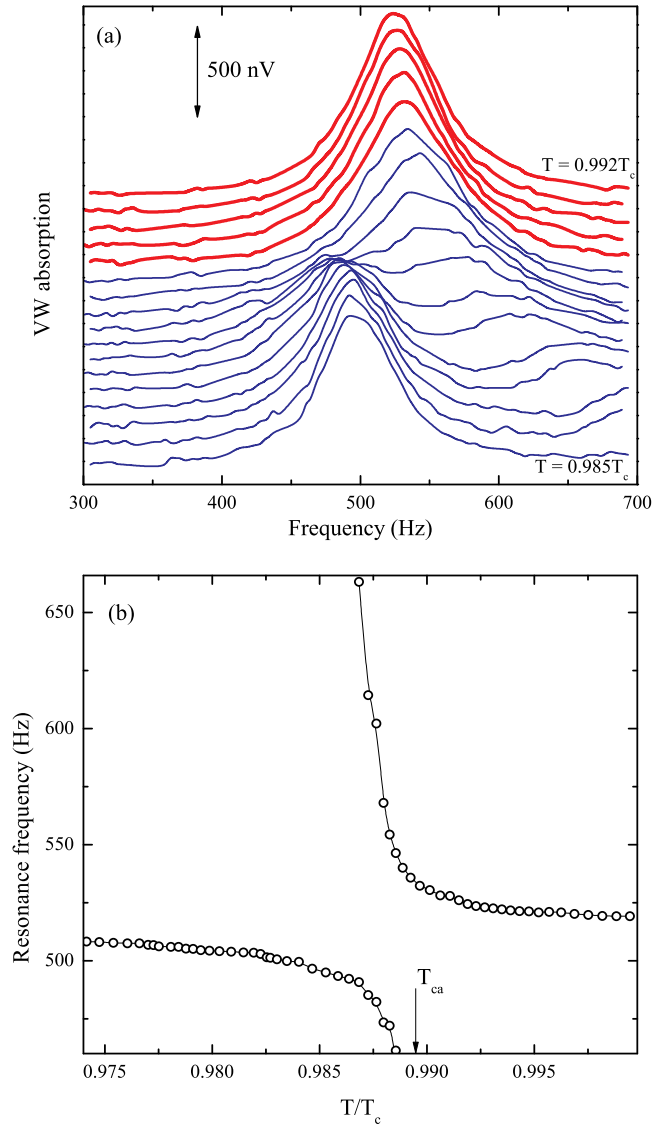}
\caption{Fig.\,\thefigure. 
(a) Temperature evolution of the VW absorption signal on slow warming from $T\approx0.985\,T_c$ to $T\approx0.992\,T_c$. For better view, the absorption lines are successively shifted upward with increasing temperature. Thick (red) and thin (blue) lines correspond to $T>T_{ca}$ and $T<T_{ca}$ respectively. (b) Two branches of the wire resonance versus temperature near $T_{ca}$ obtained by fitting the lines in panel (a) with a sum of two Lorentz peaks. $P=29.3$\,bar, $T_{ca}\approx0.989\,T_c$, $I_0=0.25$\,mA, $H=1650$\,Oe.}
\label{inter}
\end{figure}

\begin{figure}[t]
\center
\includegraphics[width=\columnwidth]{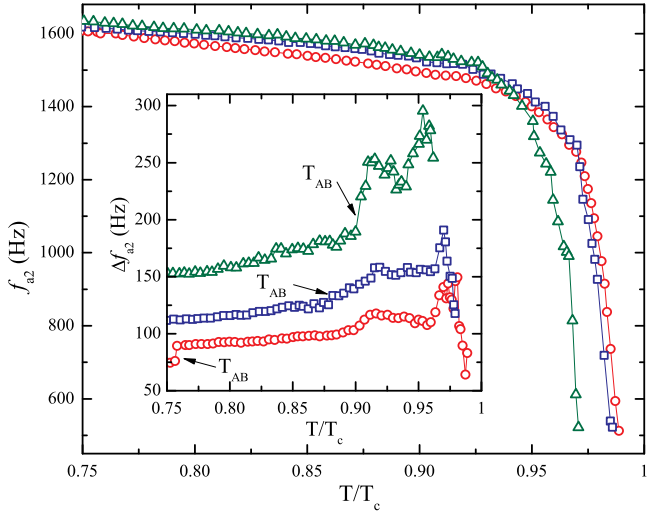}
\caption{Fig.\,\thefigure.
The resonance frequency and the resonance width (inset) of the slow sound mode in nematic aerogel versus temperature measured at $P=29.3$\,bar (circles, $T_{ca}\approx0.989\,T_c$), $P=15.4$\,bar (squares, $T_{ca}\approx0.985\,T_c$), and $P=7.1$\,bar (triangles, $T_{ca}\approx0.97\,T_c$). The given superfluid transition temperatures are nearly the same as measured in NMR experiments \cite{dmit19} with the similar sample. $I_0=0.25$\,mA, $H=1650$\,Oe.}
\label{mode2}
\end{figure}

In Fig.~\ref{mode1} we show temperature dependencies of the resonance frequency and width measured at 29.3\,bar. On cooling in normal $^3$He the resonance width is increasing and the frequency is decreasing due to the Fermi-liquid behavior of the viscosity $\propto1/T^2$ corresponding to the increase of $m_v$. Then a rapid decrease of the width (a rise of the frequency) is observed indicating a superfluid transition in bulk $^3$He at $T=T_c$. On further cooling, the second resonance appears (filled triangles in Fig.~\ref{mode1}) accompanied by the spike in the width of the main resonance.
This additional resonance mode appears just below the superfluid transition temperature of $^3$He in the sample used in NMR experiments \cite{dmit19,dmit20}. Therefore, we conclude that this resonance is due to the superfluid transition of $^3$He into the polar phase in the oscillating sample which occurs at $T_{ca}\approx0.989\,T_c$.
Although we have not been able to observe a clear resonance peak at frequencies lower than 470\,Hz, we assume that on cooling from  $T=T_{ca}$ the frequency of the second mode rapidly grows from 0 and slightly lower $T_{ca}$ becomes close to the frequency of the main resonance resulting in an interaction (repulsion) between these modes. It is illustrated by Fig.~\ref{inter}(a) where we show the evolution of the VW absorption signal during a very slow passage through $T_{ca}$. It is seen that just below $T_{ca}$ there are two resonance peaks. The temperature dependence of the resonant frequencies near $T_{ca}$ is shown in Fig.~\ref{inter}(b) that demonstrates the repulsion of two resonance modes. For clarity, below $T_{ca}$ we continue to call as the main resonance the mode with a smaller frequency. As it is seen from Fig.~\ref{mode2}, on cooling the resonance frequency of another (second) mode ($f_{a2}$) is increasing up to about 1600\,Hz at $T=0.75\,T_c$. Similar behavior of the second mode was observed at lower pressures.

We suppose that the second mode is an analog of a so-called slow sound mode observed previously in silica aerogel immersed in superfluid helium \cite{McK,gol99,Nz}. The point is that in aerogel the normal fluid component is clamped to the matrix, since $\delta$ exceeds the characteristic separation of the strands. However, the skeleton of aerogel is elastic and the normal component can move together with the strands. Therefore, the superfluid component and the combined normal fluid and aerogel matrix can move in opposite directions, resulting in a second-sound-like mode \cite{McK} which resonant frequency grows from 0 on cooling from $T_{ca}$. In superfluid $^3$He in silica aerogel such resonance mode was observed in the low-frequency sound measurements \cite{gol99,Nz}. We are dealing with a highly anisotropic aerogel which is soft in the direction normal to the strands but is rigid in the direction along the strand. Therefore, in our case the slow mode should correspond to periodic deformations of the sample in the direction normal to the strands. We note that we detect motions of the wire, but we can excite and detect the slow mode in aerogel, even if its resonance frequency is far from the original VW mechanical resonance. It means that even well below $T_{ca}$ this second resonance is strong enough to affect the wire oscillations.

\begin{figure}[t]
\center
\includegraphics[width=\columnwidth]{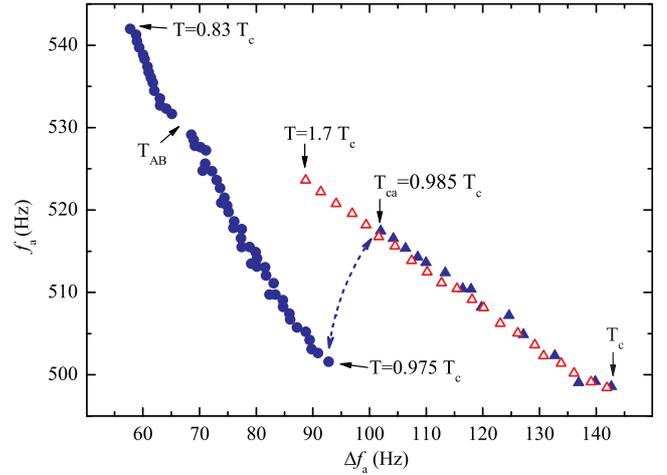}
\caption{Fig.\,\thefigure. 
The frequency of the main mode versus the resonance width measured at $P=15.4$\,bar from $0.83\,T_c$ to $1.7\,T_c$. Open triangles correspond to measurements in normal $^3$He ($T>T_c$), filled triangles are the data in the range of $T_{ca}<T<T_c$, filled circles are the data in the range of $0.83\,T_c<T<T_{ca}$. $T_{ca}\approx0.985\,T_c$. The data points in the range of $0.975\,T_c<T<T_{ca}$ where frequencies and intensities of the resonance modes are close to each other are not shown.}
\label{dens}
\end{figure}

If we neglect corrections due to that our sample is not an ellipsoid, then, the results of measurements of the frequency of the main resonance mode can be used to estimate the superfluid fraction $\rho_s^a/\rho$ of $^3$He inside the aerogel. For this purpose at $T<T_{ca}$ we can subtract contribution of $m_v$ into $f_a$, using the dependence of $f_a=f_a(\Delta f_a)$ measured at $T>T_c$. If we denote as $\tilde{f_a}$ the result of the subtraction then using Eq.~\eqref{freqs} without $m_v$ we obtain the following equation, which allows to estimate $\rho_s^a/\rho$:
\begin{equation}\label{sup}
\frac{\rho_s\rho_s^a(1+\alpha)}{\rho(\rho_s+\alpha\rho_s^a)}=1-\frac{(f_0/\tilde{f_a})^2-1}{(f_0/f_n)^2-1}.
\end{equation}
However, the existence of the second resonance makes the estimation impossible. The point is that even well below $T_{ca}$ the interaction between two resonant modes seems to be essential and their frequencies remain coupled. Fig.~\ref{dens} illustrates the influence of the second mode on the frequency of the main resonance. If we exclude data points just below $T_{ca}$ (where frequencies of resonance modes are too close to each other) then there is a jump-like decrease in the main resonance frequency due to the appearance of the second mode. At $T=0.975\,T_c$ the frequency of the main resonance is $\approx500$\,Hz and it is by 15\,Hz smaller than at $T=T_{ca}$, despite the fact that the frequency of the second mode is already much higher ($\approx1100$\,Hz). On further cooling, the frequency of the second mode continues to change, and we cannot distinguish contributions into $f_a$ from $\rho_s^a$ and from the influence of the second mode. Unfortunately, the theoretical model of the slow mode in $^3$He in aerogel described in Refs.~\cite{McK,gol99} is not applicable to our strongly anisotropic sample and further development of the theory is necessary for treatment of our results. Worthy to mark that the second order transition from the polar phase into the PdA phase should not influence the slope of the temperature dependence of $\rho_s^a$ measured along the aerogel strands \cite{sur19_2}. As it follows from the NMR experiments \cite{dmit19,dmit20}, the transition into the PdA phase in our sample should occur at $T=0.95\,T_c$ (at 29.3\,bar) and at $T=0.90\,T_c$ (at 15.4\,bar), and at these temperatures we see no any specific features in the dependencies shown in Figs.~\ref{mode1} and \ref{mode2}.

We note that Eq.~\eqref{sup} at $\rho_s=\rho$ differs from the equation used in Refs.~\cite{bru00,bru01,bra07,bra08} for determination of $\rho_s^a/\rho$ of $^3$He in silica aerogel. Using Eq.~\eqref{sup} we obtain that values of $\rho_s^a/\rho$ are about 1.3--2 times smaller (depending on temperature and $\alpha$) than that reported in Refs.~\cite{bru00,bru01,bra07,bra08}.

\section{Conclusions}
Using the aerogel wire techniques, earlier used to investigate superfluidity of $^3$He in isotropic silica aerogels, we have observed a superfluid transition of $^3$He in nematic aerogel accompanied by appearance of the second (slow sound) mode inside the aerogel sample. Resonance frequencies and widths of both the main and slow sound modes are measured in a wide range of temperatures. We think that a proper theoretical model of the slow mode in nematic aerogel might allow to estimate a superfluid density fraction inside our sample.

Our results are promising for experiments on searching for the beta phase in nematic aerogel \cite{sur19_2,sur19_1}, a new superfluid phase of $^3$He that should appear in a strong magnetic field right below $T_{ca}$. The beta phase must exist in a narrow temperature region close to $T_{ca}$ (proportional to the value of magnetic field), and on cooling from the beta phase a transition to the distorted beta phase (which is continuously transformed to the polar phase on further cooling) should be observed as a kink on a superfluid fraction versus temperature plot \cite{sur19_2}. The latter can be seen in the resonant frequencies in VW experiments. In present experiments the maximal magnetic field which we were able to apply is 1650\,Oe. In this field the range of existence of the beta phase is expected to be very small (about 0.005\,$T_c$) \cite{sur19_2}. Unfortunately, this range of temperatures is nearly the same as the range, wherein the interaction of the observed two resonance modes is rather strong and the frequency of the slow sound mode resonance is rapidly changing. This, together with experimental errors in determination of resonance frequencies, has prevented us from detecting any clear kink on temperature dependencies of $f_a$ and $f_{a2}$.

We are grateful to V.I.~Marchenko for useful discussions.

\section*{Funding}
This work was supported by the Russian Science Foundation (project no. 18-12-00384).

\end{document}